# Structures, phase transitions, and magnetic properties of Co$_3$Si from first-principles calculations

Xin Zhao[1,*], Shu Yu[1,2], Shunqing Wu[1,2], Manh Cuong Nguyen[1], Cai-Zhuang Wang[1, §], and Kai-Ming Ho[1]

[1] Ames Laboratory, US DOE and Department of Physics and Astronomy, Iowa State University, Ames, Iowa 50011, USA.

[2] Department of Physics, Xiamen University, Xiamen 361005, China.

[*]E-mail: xzhao@iastate.edu; [§]E-mail: wangcz@ameslab.gov

## Abstract

Co$_3$Si was recently reported to exhibit remarkable magnetic properties in the nanoparticle form [*Appl. Phys. Lett.* **108**, 152406 (2016)], yet better understanding of this material is to be promoted. Here we report a study on the crystal structures of Co$_3$Si using adaptive genetic algorithm, and discuss its electronic and magnetic properties from first-principles calculations. Several competing phases of Co$_3$Si have been revealed from our calculations. We show that the hexagonal Co$_3$Si structure reported in experiments has lower energy in non-magnetic state than ferromagnetic state at zero temperature. The ferromagnetic state of the hexagonal structure is dynamically unstable with imaginary phonon modes and transforms to a new orthorhombic structure, which is confirmed by our structure searches to have the lowest energy for both Co$_3$Si and Co$_3$Ge. Magnetic properties of the experimental hexagonal structure and the lowest-energy structures obtained from our structure searches are investigated in detail.

## Introduction

In recent years, considerable efforts have been devoted to discovering rare-earth (RE) free permanent magnet (PM) materials [1-9]. New RE free PM materials with enhanced magnetocrystalline anisotropy comparable to that of RE compounds [10-12] such as $Nd_2Fe_{14}B$ or $SmCo_5$ would be of great scientific and technological interest. Although such a goal is still far from being accomplished, much progress has been made both experimentally due to the development of advanced synthesis techniques [2-6] and theoretically because of the advances in the computational capability to predict new materials with promising magnetic properties [7-9].

Most of the studies so far focus on materials with positive magnetocrystalline anisotropy energies (MAE) $K_1$, while those with negative $K_1$ (easy plane anisotropy) are usually ignored. Very recently experimental study [6] showed that $Co_3Si$, exhibiting easy plane anisotropy in the bulk form, can have high coercivity and high saturation magnetization when prepared in the form of nanoparticles. This discovery draws new attentions to the materials previously overlooked.

The high temperature phase of $Co_3Si$ has been synthesized and characterized to have the hexagonal structure with space group *P*6$_3$/*mmc* (referred to as h-$Co_3Si$ structure from now on) [13-16]. Based on experimental observations, this h-$Co_3Si$ structure is only stable in a narrow temperature range of ~ 1190 to 1200 ˚C [13]. The nanoparticle phase of $Co_3Si$ was produced using non-equilibrium cluster-deposition and effective easy-axis alignment methods [6]. It was reported to have the same crystal symmetry as h-$Co_3Si$ but the lattice parameters ($a$=4.99 Å and $c$=4.50 Å) are much larger along $c$ axis than that of the bulk h-$Co_3Si$ phase ($a$=4.98 Å and $c$=4.07 Å) [15, 16], indicating the structure of the nanoparticle phase prepared at the far from equilibrium conditions is different from that of the high temperature bulk h-$Co_3Si$ phase. Using

DFT calculations [6], the h-Co$_3$Si structure ($a$=4.99 Å and $c$=4.50 Å) has been shown to exhibit an easy plane anisotropy with a high K$_1$ = –6.4 MJ/m$^3$, while the Co$_3$Si nanoparticle phase was measured to have very large uniaxial anisotropy K$_1$ = 4.8 MJ/m$^3$. These experimental and computational results are very intriguing. However, the mechanism of the structure formation and stability and the origin of strong uniaxial magnetic anisotropy in the nanoparticle phase are still not well understood.

As a first step toward a comprehensive understanding on the structure, stability and magnetic properties of the Co-Si phases formed at the non-equilibrium conditions, we investigate in this paper the energetic stability of the crystal structures of Co$_3$Si using adaptive genetic algorithm (AGA) and their magnetic properties using first-principles calculations. We show that the structure landscape of this simple binary system is much richer than what is shown in the equilibrium phase diagram [14] especially when the magnetic interactions are included. In particular, the h-Co$_3$Si phase is found to be unstable at zero temperature when magnetic interactions are included and transforms to a more stable structure with an orthorhombic symmetry which is obtained from the AGA search as the lowest-energy structure of Co$_3$Si (referred to as o-Co$_3$Si structure in this work). Other group-14 element analogs to Co$_3$Si (i.e. Co$_3$C, Co$_3$Ge and Co$_3$Sn) are also investigated.

**Computational methods**

The adaptive genetic algorithm [17, 18] used to search for the crystal structure is based on real space cut-and-paste operations [19]. AGA combines fast structure exploration by auxiliary classical potentials with accurate energy evaluation using first-principles calculations in an iterative way to ensure the efficiency and accuracy for global structure prediction. In this scheme, auxiliary classical potentials are used for the fast exploration of the configuration space, while the structures obtained from the GA search are evaluated by DFT calculations and then the results are used to guide the adjustments of the auxiliary potentials. By iteratively adjusting the potentials parameters, the adaptive GA search can explore various local basins in the configuration space effectively to locate the globally lowest-energy structure for the system with given chemical compositions.

In the AGA searches for $Co_3X$ (X = C, Si, Ge, Sn), unit cells containing up to 4 formula units are used and no symmetry constrains are imposed. During our GA search, embedded-atom method (EAM) [20] is used as the auxiliary classical potential. The potential parameters for Co-Co interaction are taken from the literature [21], while the X-X and X-Co interactions are modeled by Morse functions with 3 adjustable parameters each. The total structure population is set to be 128 in the search and convergence is considered to be reached when the lowest energy in the structure pool remains unchanged for 300 generations. In the end of each classical GA search, 16 lowest-energy structures are selected to perform first-principles calculations according to AGA procedure [17], whose energies, forces and stress are used to adjust the parameters of the EAM potential using *potfit* code [22]. Finally, all the selected structures are collected for higher-accuracy optimization using first-principles calculations.

First-principles calculations are carried out using spin-polarized density functional theory (DFT) within generalized-gradient approximation (GGA) by VASP code [23]. The GGA exchange-correlation energy functional parameterized by Perdew, Burke, and Ernzerhof (PBE) is used [24]. Plane wave basis is used with kinetic energy cutoff of 400 eV. The Monkhorst-Pack's scheme [25] is adopted for Brillouin zone sampling with a k-point grid of $2\pi$ x 0.05 Å$^{-1}$ during the AGA searches. In the final structure refinements, a denser grid of $2\pi$ x 0.03 Å$^{-1}$ is used and the ionic relaxations stop when the forces on every atom are smaller than 0.01 eV/Å. Intrinsic magnetic properties, such as magnetic moments and MAE are also calculated by VASP. Symmetry is switched off completely when performing the calculations including spin-orbit coupling and a much denser k-point grid (2π x 0.016 Å$^{-1}$) is used in the MAE calculations to achieve better k-point convergence. In this work, MAE is defined as the energy difference between aligning the spin along the easy (hard) axis and the hard (easy) plane, resulting in a positive (negative) value.

## Results and Discussions

**Phase stabilities of Co$_3$Si**

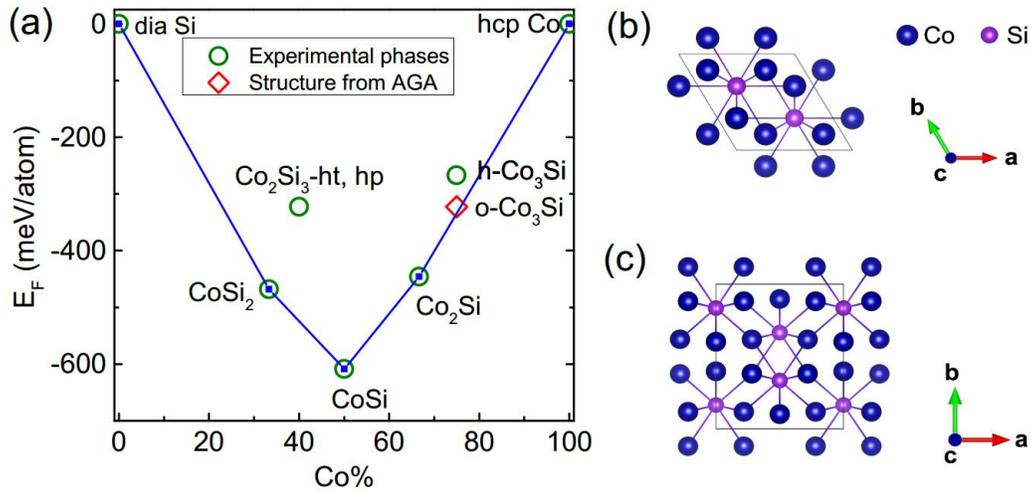

FIG. 1 (a) Formation energy convex hull of the Co-Si system. The formation energies are calculated as: $E_F(Co_cSi_{1-c}) = E(Co_cSi_{1-c}) - cE(Co) - (1-c)E(Si)$, where $0 \leq c \leq 1$ and energies of the hcp Co and diamond Si are used as references. All the phases observed in experiments are indicated by open circles (green), while the Co$_3$Si structure obtained from AGA is indicated by open diamond (red). The convex hull is constructed as the solid lines and the stable phases based on the zero temperature convex hulls are indicated by solid squares (blue). (b) The crystal structure of high temperature Co$_3$Si phase (h-Co$_3$Si) observed in experiments. (c) The new, lowest-energy structure (o-Co$_3$Si) found in our AGA searches. In (b) and (c), the bonds between Si and Co atoms are connected.

From our AGA searches of Co$_3$Si, a new orthorhombic structure with space group *Cmcm* (o-Co$_3$Si) was discovered to have the lowest energy. In Fig. 1(a), the convex hull for the Co-Si system was constructed by considering the previously reported structures: Co$_2$Si [26], CoSi [27], CoSi$_2$ [28]. It can be seen that the o-Co$_3$Si structure has lower energy than the experimentally observed h-Co$_3$Si structure with an energy difference of about 50 meV/atom. The formation energy of the o-Co$_3$Si structure is very close to (~ 9 meV above) the tie line connecting the stable Co$_2$Si and hcp Co phases in the convex hull. In the following, the stabilities of the Co$_3$Si structures will be discussed in detail.

For the h-Co$_3$Si structure, our calculation indicates the energy of its non-magnetic (NM) state is 8 meV/atom lower than that of the ferromagnetic (FM) state. Based on spin-polarized DFT relaxations, the equilibrium lattice parameters of the h-Co$_3$Si are *a*=4.97 Å and *c*=3.97 Å with the atoms occupying Co 6h (0.832, 0.664, 1/4) and Si 2c (1/3, 2/3, 1/4). While from non-magnetic calculation, a slight decrease in the volume was observed: *a*=4.94 Å and *c*=3.95 Å with the atoms occupying Co 6h (0.833, 0.667, 1/4) and Si 2c (1/3, 2/3, 1/4). The optimized *c/a* ratio (0.800) by DFT is very close to that of the bulk h-Co$_3$Si phase (0.817), but much smaller than that of the nanoparticle phase (0.902). For more details, in Fig. 2(a) we plotted the energy of the h-Co$_3$Si structure as the function of volume. The data represented by solid triangles (squares) are calculated for the FM (NM) states by fully relaxing the cell shape and atom positions while keeping the volume fixed. We can clearly see that the FM state has larger equilibrium volume and higher equilibrium energy than the NM state. It is noticed that as the decrease of volume, the FM state transforms to the NM state accompanied by a sudden drop in the Co magnetic moment as shown in the inset of Fig. 2(a). As a result, the E-V curve of the FM state (red line) jumps to overlap with that of the NM state (black line) around the volume of 9.8 Å$^3$/atom.

The E-V relation by fixing $c/a$ ratio to 0.902 according to that of the nanoparticle samples was also calculated, and the results are shown as the open symbols in Fig. 2(a). In this case, the FM state always has lower energy and smoothly transforms to the NM state when the volume is decreased. The FM to NM transition can also been clearly seen from the magnetic moment as the function of volume shown in the insert of Fig. 2(a). However, when $c/a$ is fixed as 0.902, the energy minimum is found at a much smaller volume (10.9 Å$^3$/atom) compared to the value reported for nanoparticle samples (12.1 Å$^3$/atom) [6]. In fact, when $c/a$ is fixed as 0.902, the equilibrium lattice constant is $a$=4.815 Å and $c$=4.343 Å, much smaller than those measured for the nanoparticle samples. Using the lattice constant of the Co$_3$Si phase in the nanoparticles proposed by experiment, the h-Co$_3$Si structure will give energy of 152 meV/atom higher than that of the equilibrium bulk h-Co$_3$Si phase. Although the surface effects and the synthesis procedure in the nanoparticles may contribute to stabilizing the h-Co$_3$Si structure at the lattice constant different from that of the bulk phase, it is unlikely that such effects are so strong that the lattice parameter along $c$ can be elongated by more than 10% with energy 152 meV/atom higher than that of the equilibrium bulk h-Co$_3$Si phase. The possibility that Co$_3$Si in the nanoparticle form may have a different atomic structure from the bulk h-Co$_3$Si phase cannot be ruled out. Further investigation to elucidate the atomic structure in the nanoparticle phase would be interesting.

The energy verses volume of the o-Co$_3$Si structure is plotted in Fig. 2(b). At equilibrium volume, its lattice parameters are $a$=6.26 Å, $b$=7.42 Å and $c$=3.69 Å and the atoms occupy Co 8g (0.781, 0.119, 1/4), Co 4c (0.0, 0.399, 1/4) and Si 4c (0.0, 0.834, 1/4). Unlike the h-Co$_3$Si structure, there is no spin state transition observed in the o-Co$_3$Si structure: the magnetic moment of Co atoms varies gradually with the change of volume and the non-magnetic state always has higher energy than the FM state. With the horizontal lines representing the energies of the h-Co$_3$Si structure in Fig. 2(b), we see that the energy of the o-Co$_3$Si structure is significantly lower than that of the h-Co$_3$Si structure.

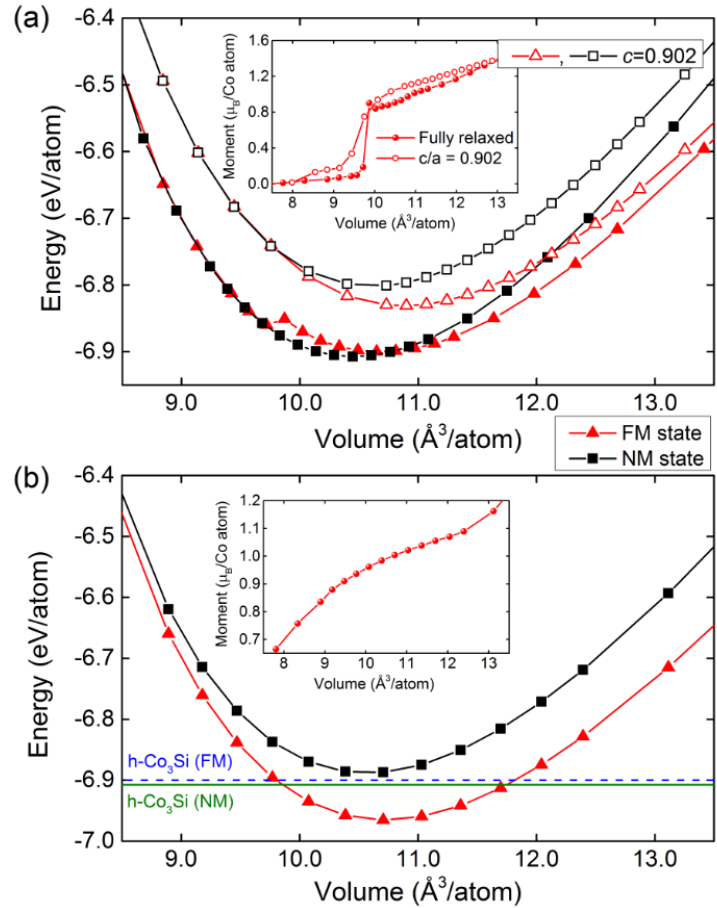

FIG. 2 Energy vs. volume plot for (a) the h-Co$_3$Si structure from both fully relaxed and fixed c/a calculations; (b) the o-Co$_3$Si structure predicted from the AGA structure searches. Open symbols

inn (a) represent data calculated by fixing the *c/a* to that of the nanoparticle samples, 0.902. The variation of the Co magnetic moments as a function of volume in both structures is plotted as the inset of (a) and (b) respectively.

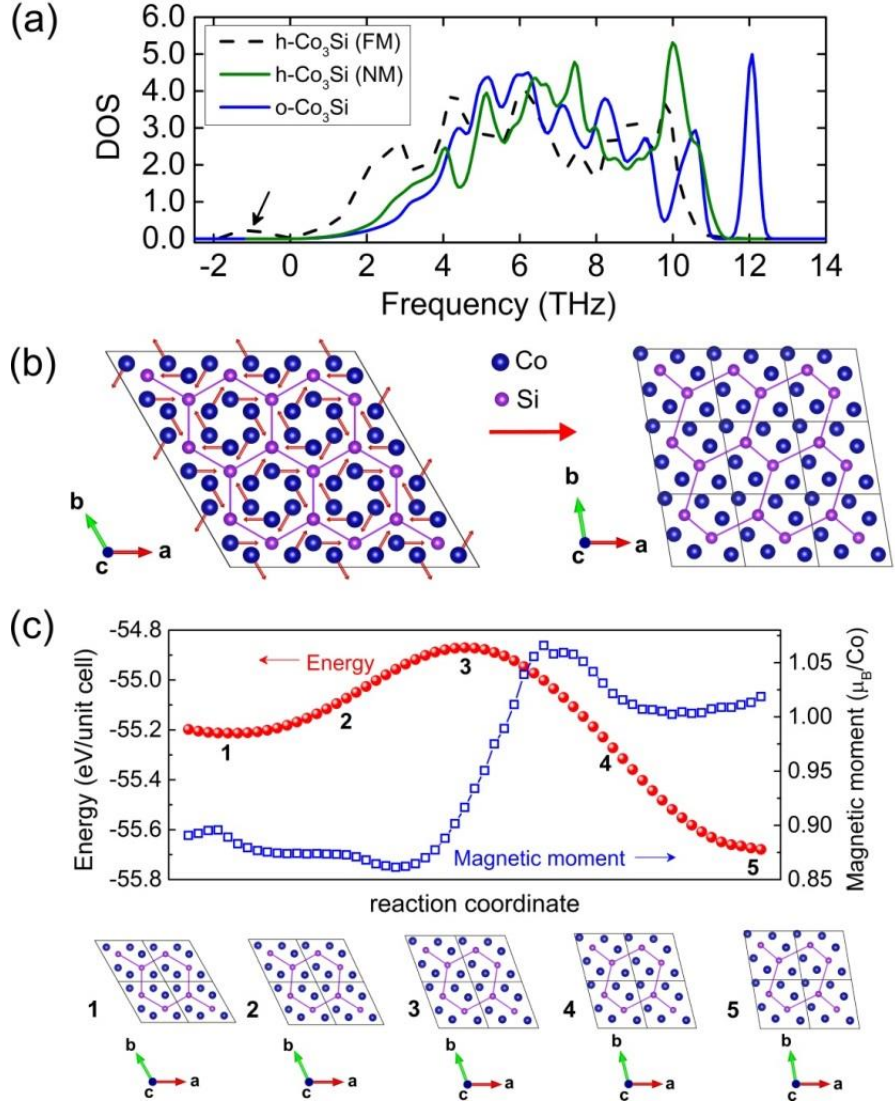

FIG. 3 (a) Phonon densities of states (DOS) calculated for the h-Co$_3$Si structure in both NM and FM states and the o-Co$_3$Si structure in FM state. Arrow points to the negative phonon frequencies in the FM h-Co$_3$Si structure. (b) Illustration of the structure transformation from the FM h-Co$_3$Si structure to the o-Co$_3$Si structure. The red arrows starting from the Co atoms in the

h-Co$_3$Si structure indicate the eigenvectors corresponding to the imaginary phonon modes. In order to accommodate the deformation with the periodic boundary condition, the unit cell of h-Co$_3$Si is expanded by 3x3x1. The o-Co$_3$Si structure plotted here is represented by its primitive cell (black boxes) with *a*=*b*=4.86 Å, *c*=3.69 Å and α=β=90°, γ=99.69°. Si-Si bonds are connected to guide the view. (c) Structure transition barrier calculated using NEB method. Both the energy and magnetic moments of Co are plotted along the transition pathway. Selected structures during the transition process are presented and labeled as 1 to 5.

We also computed the phonon properties of the h-Co$_3$Si and o-Co$_3$Si structures to investigate their dynamical stabilities. Phonon density of states (DOS) is calculated using a supercell approach provided by the *Phonopy* code [29], where supercells with sizes of 216 atoms for the h-Co$_3$Si structure and 192 atoms for the o-Co$_3$Si structure were used. The results are plotted in Fig. 3(a). There are no negative phonon frequencies in the o-Co$_3$Si structure, indicating this structure is dynamically stable. For the h-Co$_3$Si structure, it is dynamically stable in the non-magnetic state, but shows negative phonon frequencies in the FM state as indicated by the arrow in Fig. 3(a). It is known that the imaginary phonon modes can lead to new stable structures through atomic displacements and lattice deformation along the eigenvector of the soft mode [29]. We examined the eigenvector of the negative phonon mode and found that the origin of the instability comes from the Co atoms within the Si hexagon as shown in Fig. 3(b). The eigenvector corresponding to the imaginary phonon is plotted as the red arrows on the atoms in Fig. 3(b). By applying deformations following this eigenvector, the h-Co$_3$Si structure in the FM state can transform to the o-Co$_3$Si structure. The transformation pathway from the h-Co$_3$Si structure to the o-Co$_3$Si structure is shown in Fig. 3(b) and 3(c). Using the nudged elastic band (NEB) method [30], the transition barrier is estimated and the results are plotted in Fig. 3(c). It can be seen that the energy barrier for the system to move from the h-$Co_3Si$ structure to the o-$Co_3Si$ structure is about 0.34 eV/unit cell (42.5 meV/atom). Meanwhile, along the pathway of the structure transition, there is a significant increase in the magnetic moment, from ~0.88 $\mu_B$/Co atom in the h-$Co_3Si$ structure to ~1.02 $\mu_B$/Co atom in the o-$Co_3Si$ structure, although the volume is only slightly increased from 10.6 to 10.7 $Å^3$/atom.

To further understand the origin of this phase transformation, in Fig. 4, the calculated electronic DOS and band structures are plotted and compared for the h-$Co_3Si$ (FM) and the o-$Co_3Si$ structures. For both structures, the DOS near the Fermi level is mainly contributed by the 3d orbitals of Co atoms as shown in Fig. 4(a) and (b). It is also found that the Fermi level of the FM h-$Co_3Si$ structure locates at the peak position, whereas those of the o-$Co_3Si$ structures locate at the pseudogap, suggesting the phase transformation is driven by the Peierls instability [31, 32]. From the band structures plotted in Fig. 4(c) and (d), there are fewer bands in the spin down channel crossing the Fermi level of the o-$Co_3Si$ structure than those of the h-$Co_3Si$ structure, explaining the above observation in DOS.

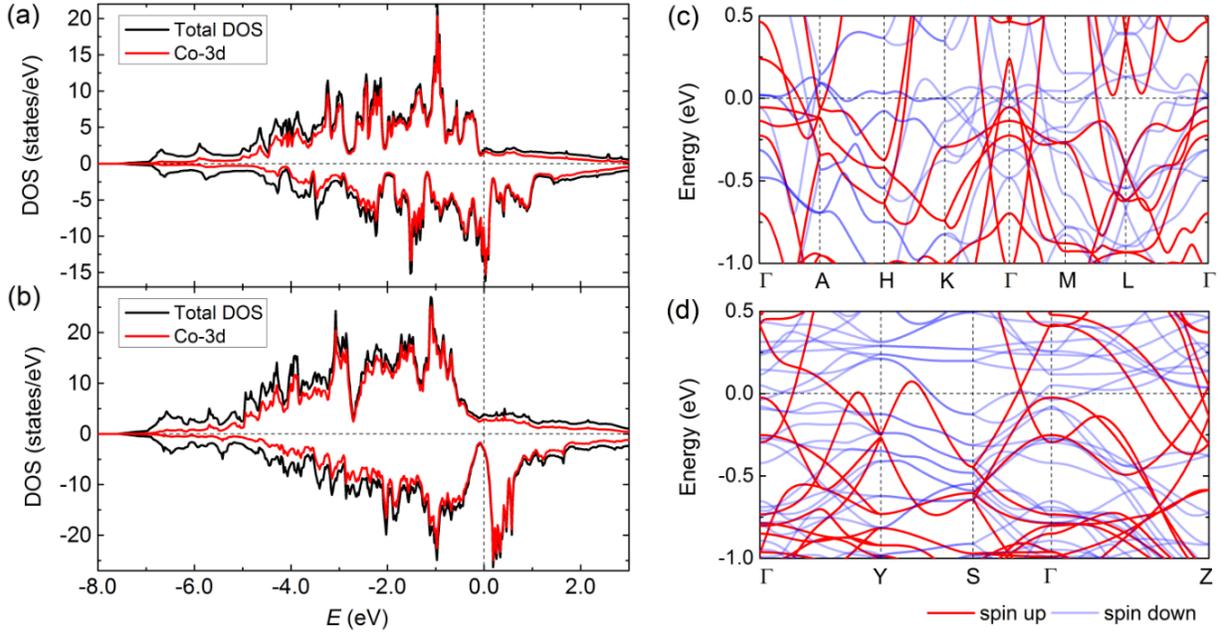

FIG. 4 (a-b) Electronic DOS of the FM h-$Co_3Si$ and o-$Co_3Si$ structures respectively. Both the total DOS and the contribution from Co 3d orbitals are presented. (c-d) Band structure of the FM h-$Co_3Si$ and o-$Co_3Si$ structures respectively along the high symmetry k-points. Fermi levels are set to zero in all plots.

Due to the fact that the h-$Co_3Si$ bulk phase can only be stabilized in a narrow high temperature range and was reported to decompose to the Co and $Co_2Si$ at lower temperatures, until now there is no experimental studies on the magnetic properties of bulk $Co_3Si$ samples. Future experiments to verify our prediction of the magnetic state of bulk h-$Co_3Si$ structure at low temperatures as well as the h-$Co_3Si$ to o-$Co_3Si$ structure transition would be very interesting. Based on our calculation at zero temperature, the new o-$Co_3Si$ structure wins over the h-$Co_3Si$ structure by more than 50 meV/atom and the energy barrier for the proposed structure transition is estimated to be about 42.5 meV/atom, which roughly corresponds to a temperature difference of 500 K.

Therefore the o-$Co_3Si$ phase should be very robust and have high probability to be observed in experiment.

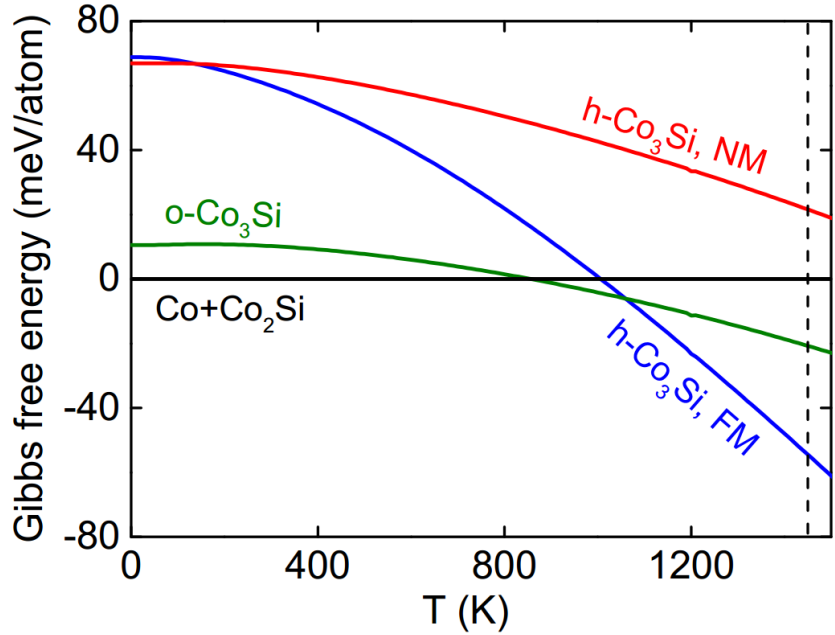

Fig. 5 Gibbs free energy of the $Co_3Si$ structures relative to the decomposition reaction Co + $Co_2Si$ calculated from quasi-harmonic approximations. The dash vertical line represents the melting temperature at the composition of $Co_3Si$ [14].

In order to gain some insights into the effect of temperature, we used quasi-harmonic approximations to estimate the changes of the Gibbs free energies as the function of temperature for the different phases of $Co_3Si$ with respect to Co+$Co_2Si$ decomposition using Phonopy code [29]. Calculations were performed for all three $Co_3Si$ phases, hcp Co and $Co_2Si$ at 11 different volumes to get the Helmholtz free energy as function of the unit cell volume. Temperature effect is taken into account via vibrational phonon entropy, where the imaginary phonon modes of h-$Co_3Si$ are excluded. In addition, we note that in our calculations the magnetic moments of Co atoms are kept at zero temperature values, i.e. the magnetic phase transition was not considered automatically as temperature changes.

The results are plotted in Fig. 5. It can be seen that as the temperature increases, the free energies of NM h-$Co_3Si$ and o-$Co_3Si$ structures decrease much slower than that of the FM h-$Co_3Si$ structure, because the FM h-$Co_3Si$ structure has more low-frequency phonon modes as shown in Fig. 3(a). The free energy of NM h-$Co_3Si$ is always above other phases. Above ~850 K, the o-$Co_3Si$ structure becomes the most stable phase for a narrow temperature window. The FM h-$Co_3Si$ phase is estimated to have the lowest free energy above ~1050 K, which is quite close to the melting temperature (~1450 K) at the composition of $Co_3Si$ [14]. Note the transition temperature here can be different from the experimental observation due to the approximations used in the free energy calculation. Despite of this, the free energy plot above provides some useful insights for understanding the trends of phase stability at finite temperatures and could serve as a reference to the experimental synthesis effort.

**Group-14 analogs of $Co_3Si$**

Due to the chemical similarities among C, Si, Ge and even Sn, it is interesting to look at the same stoichiometric compound of Co and other group-14 elements. Our crystal structures searches were also performed for $Co_3X$ with X = C, Ge and Sn. For C or Sn, no structures were found to have negative formation energies with respect to hcp Co and pure C or Sn diamond structure, thus $Co_3C$ and $Co_3Sn$ will not be discussed in the present work.

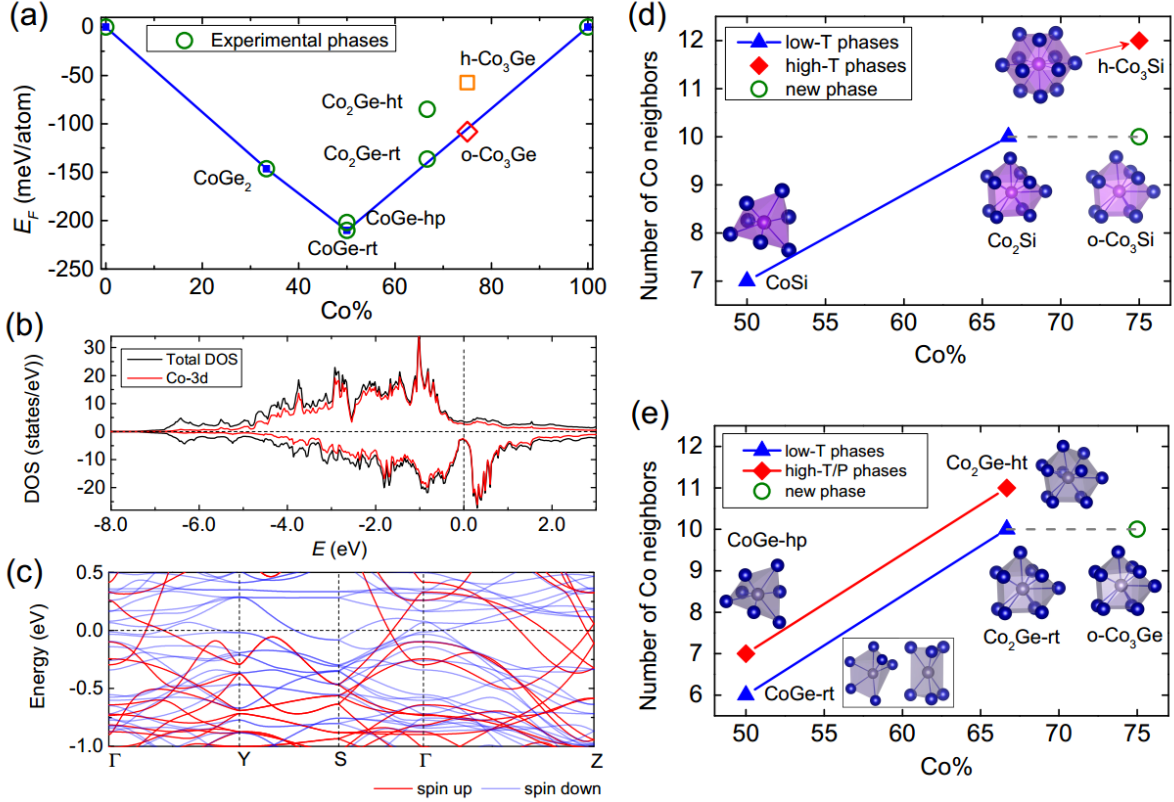

FIG. 6 (a) Formation energy convex hull of the Co-Ge system. Energies of the hcp Co and diamond Ge are used as references to calculate the formation energies. The crystal structures of the experimental phases indicated by open circles come from Ref. [33]. (b, c) Electronic DOS and band structure calculated for the o-$Co_3Ge$ structure. (d, e) Number of neighboring Co atoms connected to Si/Ge as the function of Co composition in the known structures of Co-Si and Co-Ge systems. Triangles (blue) represent the low temperature phases and diamonds (red) represent the high temperature or high pressure phases. The new structure discovered in this work (o-$Co_3X$) is indicated by the open circles (green). Among all the considered structures, CoGe-rt structure has two types of Ge-centered clusters, while only one type of Si/Ge-centered cluster exists in the other structures because of their symmetries.

It has been reported that Co$_3$Ge could exist as a stable phase at 700 and 750 °C, but its crystal structure remains unclear and has been speculated to be the A15-type structure [14]. Our structure search demonstrated that the lowest energy structure of Co$_3$Ge is the same as o-Co$_3$Si (referred to as o-Co$_3$Ge from now on). At equilibrium volume, its lattice parameters are $a$=6.43 Å, $b$=7.55 Å and $c$=3.75 Å and the atoms occupy Co 8g (0.224, 0.116, 1/4), Co 4c (0.0, 0.400, 1/4) and Ge 4c (0.0, 0.832, 1/4). In comparison, the formation energy of the A15-type structure is positive and ~125 meV/atom higher than that of the o-Co$_3$Ge structure. Furthermore, the hypothetical h-Co$_3$Ge structure was relaxed and computed by replacing Si with Ge in the h-Co$_3$Si structure, but its energy in the FM state is about 51 meV/atom higher than that of the o-Co$_3$Ge structure. The convex hull of the Co-Ge system is constructed in Fig. 6(a) and the o-Co$_3$Ge structure lies below the line connecting the CoGe phase and hcp Co (~ 3 meV/atom below). Thus, the crystal phase observed in experiment [14] is more likely to be the o-Co$_3$Ge structure rather than the A15-type structure. Nonetheless, given that the formation energy of the room temperature Co$_2$Ge phase is 6.8 meV/atom above the convex hull and it has been successfully synthesized, the experimental realization of o-Co$_3$Ge would be very possible.

Given the similarities between Si and Ge, the electronic properties of o-Co$_3$Si and o-Co$_3$Ge are also anticipated to be similar. In Fig. 6(b) and (c), the electronic DOS and band structure are plotted for the o-Co$_3$Ge structure. It can be clearly seen that both the DOS and band structure of the o-Co$_3$Si (plotted in Fig. 4(b) and (d)) and o-Co$_3$Ge are very alike. The Fermi level of the o-Co$_3$Ge structure locates right at the valley around the pseudogap, indicating the stable nature of the new structure and our phonon calculation also supports its dynamical stability by showing no imaginary frequencies.

To characterize the structures for different Co compositions in Co-Si and Co-Ge systems and understand the connections between them, we analyzed the local environment of the Si/Ge atoms in the previously reported and the newly found structures. The crystal structures of CoX, $Co_2X$ (X=Si and Ge) are taken from the experimental studies in the literature [33]. It can already be seen from Fig. 1(b) and (c) that the o-$Co_3$X structure has very different bonding character compared to the h-$Co_3$Si structure. The h-$Co_3$Si structure can be considered as Si-substituted hcp Co, i.e. each Si atom bonds with 12 neighboring Co atoms, while in the new structure, each Si or Ge atom bonds with 10 Co atoms, forming a distorted equatorial four-capped trigonal prism. To find the connection between the different phases, in Fig. 6(d) and (e), the number of neighboring Co atoms connected to Si/Ge in Co-Si and Co-Ge systems was plotted as a function the Co composition. As the Co composition is increased, more Co atoms are bonded with Si/Ge as expected. We also note Si/Ge atoms in the structures observed at high temperature or high pressures tend to have more Co neighbors. In addition, it is of particularly interest to see that the new o-$Co_3$X structure is closely related to the stable $Co_2$X phases: the same Si/Ge centered cluster (X$Co_{10}$) is the building block for both the low temperature $Co_2$X phases and the newly discovered o-$Co_3$X structures. In fact, there is only one type of Si/Ge centered cluster in these structures and the different arrangements and orientations of such X$Co_{10}$ clusters, i.e. having different numbers of face or vertex sharing, gives different Co compositions.

**Magnetic properties**

As mentioned above, the $Co_3$Si structure synthesized in the nanoparticle form shows remarkable magnetic properties. Although the structure of this nanoparticle phase is claimed to have the same crystal symmetry as the high-temperature bulk h-$Co_3$Si from XRD analysis [6], the lattice parameters and volume are very different. The volume of the bulk h-$Co_3$Si structure was reported to be ~ 10.9 Å$^3$/atom [15, 16], which is rather close to the equilibrium volumes calculated by spin-polarized GGA (10.6 Å$^3$/atom). In the Co$_3$Si nanoparticles [6], the volume was reported to be 12.1 Å$^3$/atom. To gain a comprehensive understanding of the magnetic properties of the h-Co$_3$Si structure, we calculated the MAE and magnetic moments using the atomic structure from the bulk h-Co$_3$Si phase but varying the lattice parameters $a$ and $c$. The results are summarized as contour plots in Fig. 7. First we note again that as $a$ varies from 4.5 Å to 5.5 Å and $c$ varies from 3.5 Å to 5.0 Å, there is only one energy minimum at the lattice parameters close to those of the bulk h-Co$_3$Si phase, as plotted in Fig. 7(a).

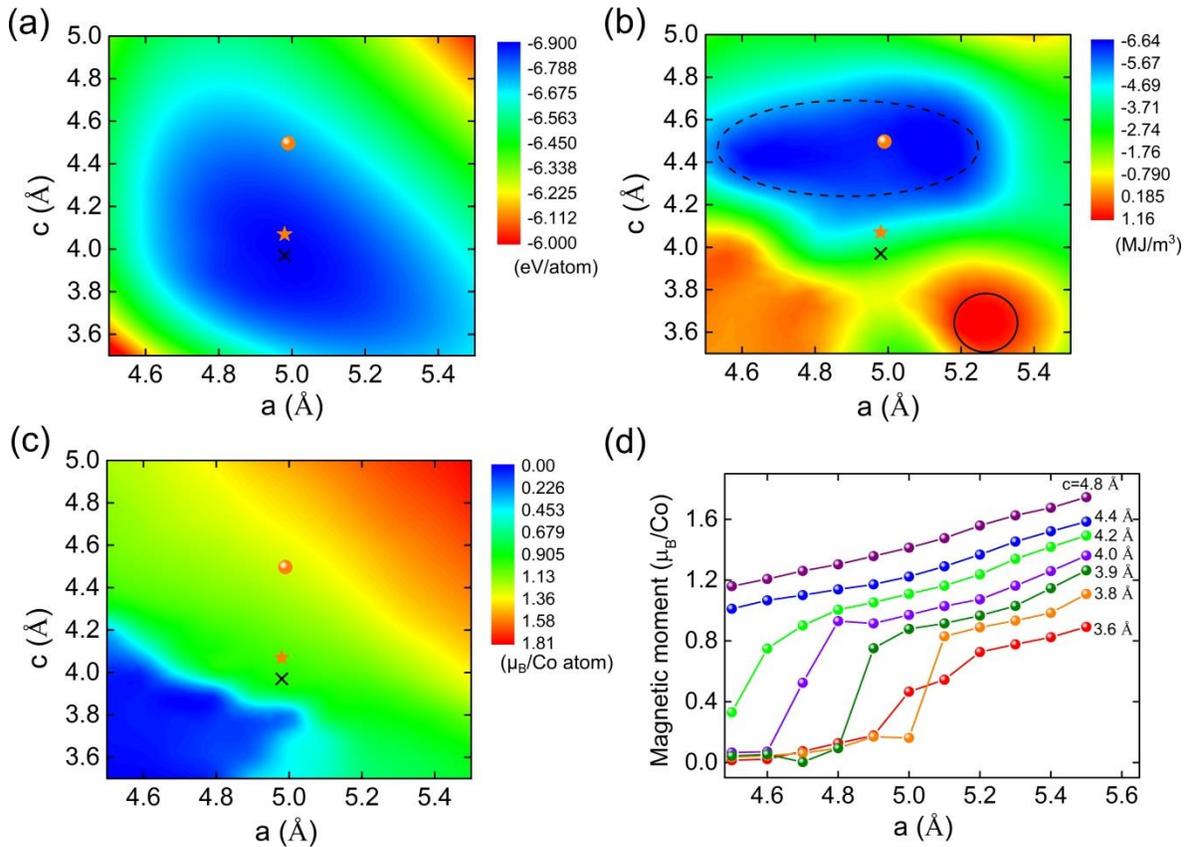

FIG. 7 Contour maps of the (a) energy, (b) magnetic anisotropy energy and (c) Co magnetic moment of the h-Co$_3$Si structure as functions of its lattice parameters $a$ and $c$. The star and ball represent the lattice parameters of the experimentally observed bulk sample and nanoparticle

sample respectively. Black cross represents the position of the DFT optimized lattice parameters. In (b), the dashed circle (black) highlights the areas with large easy plane magnetic anisotropy, while solid circle (black) highlights the areas with uniaxial magnetic anisotropy. (d) The variation of the Co magnetic moment as a function of $a$ for fixed $c$'s.

From Fig. 7(b), we can see that MAE varies dramatically with the lattice parameters $a$ and $c$. Using $a$ and $c$ of the nanoparticle samples, our calculation shows a MAE value of ~ –6.1 MJ/m$^3$ (–0.46 meV/atom), consistent with the previous calculation [6]. This result contradicts to the experimental measurement as has been discussed in Ref. [6], also suggesting that the crystal structure of the nanoparticle phase may be different from the bulk h-Co$_3$Si phase. When using the DFT optimized structure or the lattice parameters from bulk samples, the calculated MAE is much smaller, ~ –2.7 MJ/m$^3$ and –3.5 MJ/m$^3$ respectively. In fact, three distinct regions can be clearly identified in Fig. 7(b): values of $a$ and $c$ in the left bottom corner give no magnetic anisotropy, which can be attributed to the zero magnetic moment as shown in Fig. 7(c). Values of $a$ and $c$ in right bottom corner as circled by the solid black line exhibit uniaxial magnetic anisotropy (~ 1 MJ/m$^3$), and those in the top region as circled by the dashed black line have an easy plane magnetic anisotropy that goes as large as –6.7 MJ/m$^3$. Here it is demonstrated again that changing the lattice parameters of the crystal structure can significantly affect the MAE although the space group symmetry of the crystal is fixed [34-36]. The variations of magnetic moments with $a$ and $c$ are plotted in Fig. 7(c) and (d). In general, the magnetic moment increases gradually as the unit cell volume increases. Meanwhile, we also observed from Fig. 7(d) that for $c$ smaller than 4.2 Å, there exists a sudden rise in the magnetic moment as $a$ increases, but for c = 4.5 Å as in the nanoparticle samples, such phenomenon does not occur.

As for the lowest-energy o-Co$_3$Si structure, the calculation shows that it has an easy plane magnetic anisotropy with MAE equal to –0.52 MJ/m$^3$ and the magnetic hard axis is along its *a* direction. On the other hand, the o-Co$_3$Ge structure, from our magnetic calculation, shows uniaxial magnetic anisotropy with the easy axis along its *c* direction, but its MAE is very small, ~ 0.35 MJ/m$^3$.

**Conclusion**

Our first-principles calculations on the experimentally reported hexagonal Co$_3$Si structure reveal a spin state transition that has not yet been observed in experiments. At zero temperature, the h-Co$_3$Si structure has lower energy in the NM state than the FM state. The h-Co$_3$Si structure in the FM state is dynamically unstable based on phonon calculations. With the assistance of adaptive genetic algorithm crystal structure searches, the lowest-energy structure of Co$_3$Si is found to be the o-Co$_3$Si structure with energy ~ 50 meV/atom lower than that of the h-Co$_3$Si structure. A structure transformation from the ferromagnetic h-Co$_3$Si structure to the o-Co$_3$Si structure is then demonstrated and the transition barrier is estimated to be 42 meV. Though quasi-harmonic approximations, we discussed the structure evolution of the different Co$_3$Si phases as the function of temperatures and predict the possible temperature window to obtain the new structures. By extending the study to other group 14 analogs, we show the lowest-energy structure of Co$_3$Ge is the same as o-Co$_3$Si structure. This o-Co$_3$X structure is closely related to the stable Co$_2$X phases in terms of the constituent structural motifs.

Magnetic properties are discussed for both the experimental and new structures of Co$_3$X. In order to address the effect of the different lattice parameters as observed for bulk and nanoparticle Co$_3$Si samples on the magnetic properties, a comprehensive dependence of MAE and magnetic moments on the lattice parameters is mapped out and discussed. For the new structure, although it is found to have lower energies in both the Co-Si and Co-Ge systems, the calculated magnetic anisotropy energies of them are very small. Nonetheless, the proposed structure and spin transitions from our present study would provide a strong motivation for devoting further experimental efforts to explore the formation of new phases under far from equilibrium conditions in the Co-Si and Co-Ge systems.

## Acknowledgment

This work was supported by the National Science Foundation (NSF), Division of Materials Research (DMR) under Award DMREF: SusChEM 1436386. The development of adaptive genetic algorithm (AGA) method was supported by the US Department of Energy, Basic Energy Sciences, Division of Materials Science and Engineering, under Contract No. DE-AC02-07CH11358, including a grant of computer time at the National Energy Research Scientific Computing Center (NERSC) in Berkeley, CA. SY also acknowledges the support from China Scholarship Council for visiting Iowa State University under the File No. 201506310134.